\documentclass[journal]{IEEEtran}

\ifCLASSINFOpdf
\else
   \usepackage[dvips]{graphicx}
\fi
\usepackage{url}
\usepackage{amsmath}
\usepackage{amsfonts}
\usepackage{graphicx}
\usepackage{algorithm}
\usepackage[noend]{algpseudocode}
\usepackage{cite}

\begin{document}

\title{Core consistency diagnosis for Block Term Decomposition in rank ($L_r, L_r, 1$)}

\author{Noramon Dron$^{1}$, \IEEEmembership{Member, IEEE}, Javier Escudero$^{2}$, \IEEEmembership{Senior Member, IEEE}
\thanks{*This work was supported by a PhD studentship to N.~Dron by the Royal Thai Government Scholarship. }
\thanks{$^{1}$N.~Dron is with the School of Engineering, Institute for Imaging, Data and Communications, University of Edinburgh, Scotland, UK and the Department of Biomedical Engineering, Srinakharinwirot University, Thailand.
 {\tt\small noramon.info@gmail.com}}
\thanks{$^{2}$J.~Escudero is with the School of Engineering, Institute for Imaging, Data and Communications, and with the Muir Maxwell Epilepsy Centre, University of Edinburgh, Scotland, UK}}

\maketitle

\begin{abstract}
Determining the underlying number of components $R$ in tensor decompositions is challenging. Diverse techniques exist for various decompositions, notably the core consistency diagnostic (CORCONDIA) for Canonical Polyadic Decomposition (CPD). Here, we propose a model that intuitively adapts CORCONDIA for rank estimation in Block Term Decomposition (BTD) of rank $(L_r, L_r, 1)$: BTDCORCONDIA. Our metric was tested on simulated and real-world tensor data, including assessments of its sensitivity to noise and the indeterminacy of BTD $(L_r, L_r, 1)$. We found that selecting appropriate $R$ and $L_r$ led to core consistency reaching or close to 100\%, and BTDCORCONDIA is efficient when the tensor has significantly more elements than the core. Our results confirm that CORCONDIA can be extended to BTD $(L_r, L_r, 1)$, and the resulting metric can assist in the process of determining the number of components in this tensor factorisation.
\end{abstract}

\vspace{-0.5em}

\IEEEpeerreviewmaketitle

\section{Introduction}

\IEEEPARstart{T}{ensor} decompositions are important techniques for handling and analysing multidimensional data. Tensor decomposition has been used in many fields of research, such as feature extraction \cite{phan2010tensor,kisil2018common}, signal processing \cite{Cichocki2013TensorAnalysis,cichocki2015tensor,sidiropoulos2017tensor}, and many biomedical problems \cite{Cong2015TensorReview,acar2017tensor,Hunyadi2017TensorData,chatzichristos2020early,van2021augmenting}. However, it often lacks ground truth about the true number of components to estimate \cite{bro2003new,Kolda2009TensorApplications}. Exploring all the combinations of the number of components is time-consuming. In this context, the idea of rank estimation and core consistency diagnosis (CORCONDIA) was introduced \cite{bro2003new} to assist the estimation of $R$ in canonical polyadic decomposition (CPD).

Following \cite{bro2003new}, several works have attempted to improve the CORCONDIA for CPD in various aspects, such as improved computation performance for big sparse tensors \cite{papalexakis2015fast}, for compressed tensors \cite{tsitsikas2019core}, and for noise robust CPD \cite{das2020efficient}. Additionally, several works adopted the idea and extended it to cover Tucker3 \cite{kompany2012tucker}, and PARAFAC2 \cite{kamstrup2013core}. 

Block term decomposition (BTD), especially the BTD $(L_r, L_r, 1)$, is a tensor decomposition that has recently risen in popularity, such as in various biomedical applications \cite{Hunyadi2014BlockSeizures,chatzichristos2017higher,chatzichristos2019blind,de2021blind}.  
BTD $(L_r, L_r, 1)$ can be considered a generalised version of CPD, with two modes varying while one stays at rank 1. Since CPD has a core identity tensor \cite{bro2003new}, BTD can be seen as having a similar core tensor composed of patterns of 1 and 0 as well. In CPD, CORCONDIA can estimate the number of components $R$ \cite{bro2003new} by comparing the elements in the core to how close they are to the ideal identity tensor. Similarly, we hypothesise that a similar diagnostic model can be built specifically for BTD $(L_r, L_r, 1)$ to estimate the number of components $R$ and rank $L_r$. 

Thus, our main contributions include investigating the BTD $(L_r, L_r, 1)$ core tensor and proposing a core consistency diagnosis, so-called BTDCORCONDIA, to guide the determination of the number of components to extract and decide if a tensor follows BTD $(L_r, L_r, 1)$. We explore the performance of BTDCORCONDIA with simulated and real-world data, as well as noise sensitivity and indeterminacy of BTD $(L_r, L_r, 1)$.
 



\vspace{-0.5em}
\section{Background}

\subsection{CPD and its CORCONDIA}
Without loss of generality, suppose we have a three-way tensor $\mathcal{X}$. CPD can decompose it into a sum of rank-one vectors as:
\begin{equation}\label{eq1}
\mathcal{X}  \approx  \sum_{r=1}^{R} \textbf{a}_r\circ \textbf{b}_r \circ \textbf{c}_r,
\end{equation}
where $\circ$ denoted to outer product, $r = 1, 2, \ldots, R$ with $\textbf{a}_r\in\mathbb{R}^I,\:\textbf{b}_r\in\mathbb{R}^J,\: \textbf{c}_r\in\mathbb{R}^K$. Those vectors then can be grouped into matrices $\textbf{A}\in \mathbb{R}^{I\times R}$, $\textbf{B}\in\mathbb{R}^{J\times R}$,  and $\textbf{C}\in \mathbb{R}^{K\times R}$ so that CPD  can alternatively be written as
\begin{equation}\label{eq1}
\mathcal{X} \approx \mathcal{I} \times_1 \textbf{A} \times_2 \textbf{B} \times_3 \textbf{C},
\end{equation}
where $\times_n$ denotes tensor-matrix product in mode $n$ and $\mathcal{I}$ is a core identity tensor where position $i=j=k$ has value $1$ and $0$ elsewhere. For an ideal CPD, its core tensor will always be an identity tensor $\mathcal{I}$. This led to the concept of CORCONDIA \cite{bro2003new}, defined as
\begin{equation}\label{eq_corcondia}
\mathrm{CORCONDIA} = \left(1 - \left( \frac{|| \mathcal{I} - \mathcal{G} ||^2}{||\mathcal{I}||^2}  \right) \right) \times 100.
\end{equation}

CORCONDIA can help determine the number of components $R$ for the CPD by assessing the interaction between modes \textbf{A},  \textbf{B}, and \textbf{C} to the tensor $\mathcal{X}$ \cite{bro2003new}. If the number of components $R$ is suitable for the model, and the interactions are appropriate, the core tensor $\mathcal{G}$ should closely resemble $\mathcal{I}$ and yield a value close to $100 \%$ in CORCONDIA. CORCONDIA can straightforwardly be extended to the $n$-way tensor, making this a powerful approach for CPD.

\subsection{Block Term Decomposition ($L_r, L_r, 1$)}
BTD ($L_r, L_r, 1$) or BTD for short \cite{de2008decompositions,de2008decompositions2}, is a generalised version of CPD where two modes are varied together while the other mode is fixed at rank 1. For a 3-way tensor, BTD can be expressed as
\begin{equation}\label{BTD_eq1}
\mathcal{X}  \approx  \sum_{r=1}^{R} (\textbf{A}_r \cdot \textbf{B}_r^\top) \circ \textbf{c}_r
\end{equation}
where factor matrices $\textbf{A}_r \in \mathbb{R}^{I \times L_r},\:\textbf{B}_r\in\mathbb{R}^{J \times L_r}$, $\textbf{c}_r\in\mathbb{R}^K$, and $1 \leq r \leq R$. 
This can be seen as CPD when written as 
\begin{equation}
\textbf{X} \approx \sum^R_{r=1}\sum^{L_r}_{l=1}(\textbf{A}_r(:,l) \otimes \textbf{B}_r(:,l) \otimes \textbf{c}_r,
\end{equation}

where $\otimes$ denoted to Kronecker product. Alternatively, the equation can also be expressed in Tucker form as



\begin{equation}\label{BTDcore_eq}
\mathcal{X}  \approx  \mathcal{G} \times_1 \textbf{A} \times_2 \textbf{B} \times_3 \textbf{C}.
\end{equation}


\section{Method}

\subsection{Core inspection}

\begin{figure}[!ht]
      \centering
      \includegraphics[width=0.8\linewidth]{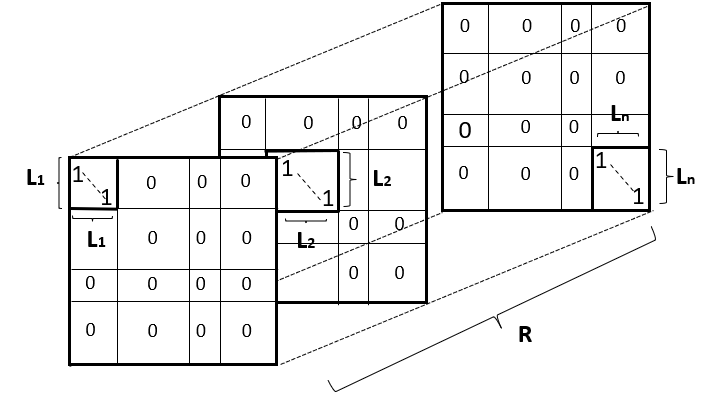}
      \caption{The ideal core tensor $\mathcal{G}$ for BTD $(L_r, L_r, 1)$. The core always consists of slices of identity matrix \textbf{I} of size $L_r \times L_r$, diagonally stacking in a total of $R$ slices.}
      \label{BTDcorcond_idealcore}
      \vspace{-0.5em}
\end{figure}

When all $L_r=1$, BTD reduces to CPD. Hence, in ideal circumstances, the core tensor $\mathcal{G}$ would equal $\mathcal{I}$. Building on this idea, since the factors interact within the same component $R$ as in the CPD, the ideal core tensor $\mathcal{G}$ of the BTD must have similar variations of the identity values of only 1 and 0. 
Thus, the ideal core $\mathcal{G}$ will always consist of slices of identity matrix \textbf{I} of size $L_r \times L_r$, diagonally stacking in a total of $R$ slices while the rest of the tensor are zeros as shown in Fig.~\ref{BTDcorcond_idealcore}. 

\subsection{Proposed BTDCORCONDIA}
We extend the CORCONDIA concept \cite{bro2003new,kompany2012tucker} to work with the BTD. The elements inside the ideal core tensor $\mathcal{G}$ will vary according to the number of $L_r$ and $R$, but they will always be 0 and 1. We name this ideal core for BTD as $\mathcal{II}$.

The core tensor $\mathcal{G}$ can be calculated from the original tensor and its factor matrices. Therefore, the BTD core consistency diagnostic (BTDCORCONDIA) can be defined  as
\begin{equation}\label{BTD_eq_corcond}
\mathrm{BTDCORCONDIA} = \left(1 - \left( \frac{|| \mathcal{II} - \mathcal{G} ||^2}{||\mathcal{II}||^2}  \right) \right)\times 100.
\end{equation}


BTDCORCONDIA close to 100\% indicates that the tensor is suitable for BTD. The consistency value can be negative, and it is capped at  100\%.

The BTDCORCONDIA heuristic is built using MATLAB and is available on: \url{https://github.com/NoramonDron/BTDCORCONDIA}. The method takes a tensor, factors estimated with BTD, and $L_r$ as input. The model calculates the core tensor $\mathcal{G}$ and generates the ideal core tensor $\mathcal{II}$ according to the $L_r$ and $R$. (The model extracts the value of $R$ from the input $L_r$.) Lastly, the output core consistency percentage is calculated with equation~\ref{BTD_eq_corcond}. 

            
        

\subsection{Experiments}

\subsubsection{Simulated data}\label{corcond_simdata}

To test BTDCORCONDIA, tensors of size $50 \times 60 \times 70$ are created from the combination of $L_r$ and $R$. $L_r$ values ranged from 2 to 6, and $R$ values between 1 and 6 were tested. Tensors are divided into two groups: without noise and with added noise. Noise is added with a 50 dB signal-to-noise ratio (SNR).

The generated tensors went through conventional BTD from Tensorlab 3.0 toolbox (ll1 algorithm) \cite{Vervliet2017TensorlabFactorization}, with generalised eigenvector decomposition (gevd) initialisation and without other constraints to obtain the factors. The gevd initialisation is suggested in \cite{de2008decompositions2} to ensure the BTD could reach the global solution. 

A range of values of $R$ and $L_r$ were used to estimate BTD, and the values of BTDCORCONDIA were computed. We investigated whether the model could accurately find the actual $R$ and $L_r$, ideally should result in $\approx100 \%$ core consistency. Each $L_r$ and $R$ combination will be repeatedly run 100 times, with and without noise, to investigate the core consistency percentage's mean and standard deviation (SD).



 
 
 
 
 
 
 
 
 
 
 
 
 
 
 
 




To inspect the model's noise sensitivity, increasing noise levels were added to the factors extracted from the ideal tensor, and then the core consistency was plotted against the SNR.

\subsubsection{Exploration of the BTD indeterminacy} 

It is known that, in the original CORCONDIA, the number of elements in the tensor should significantly exceed the number of elements inside the core $\mathcal{G}$ so that the core is stable. Thus, CORCONDIA should be avoided for very small tensors \cite{bro2003new}. To determine if the same holds for BTDCORCONDIA, small tensors of size $10 \times 11 \times 12$ were tested. The tensor underwent the same process as the larger tensor in Section~\ref{corcond_simdata}.

Additionally, one of the properties of BTD is the arbitrary permutation, where one can postmultiply any non-singular matrix $\textbf{F}_r$ size $L_r\times L_r$ to factor $\textbf{A}_r$, provide that factor $\textbf{B}_r$ is also premultiplied with the inverse of $\textbf{F}_r$ \cite{de2008decompositions}. 
Therefore,  the tensor was generated from different combinations of $L_r$ and $R$ and then went through BTD with the same $L_r$ and $R$ to obtain the factors. A random matrix $\textbf{F}_r$ and its inverse were generated and applied to the factor components. The original factors and factors with apply matrix $\textbf{F}_r$ go through the BTDCORCONDIA model separately. Then, the core consistency from both cases was compared. 

\subsubsection{Real data}
Real data sets are put to the BTDCORCONDIA test.
We opt for the electroencephalogram (EEG) tensors derived from the NEUROPROFILE (Neu) dataset \cite{Hunter2015Neurodevelopment099} for epilepsy subjects and the Child Mind Institute - Healthy Brain Network (CMI) \cite{alexander2017open} for healthy subject. Both data are processed and put in tensor form following the work in \cite{dron2021functional}. Neu and CMI EEG tensors have modes of subject by frequency by channels, with tensor sizes of $12 \times 79 \times 20$ and $24 \times 81 \times 111$, respectively. Both tensors go through BTD with gevd initialisation and without additional constraints. Then, we explore all the combinations of $R$ and $L_r$, from 1 to 6, to find the best fit. After that, the EEG tensor, its factors and $L_R$ are put into BTDCORCONDIA to calculate the core consistency. The values of $R$ and $L_r$ that give the highest core consistency percentage were presented and analysed.

\vspace{-0.5em}
\section{Results}\label{corcond_results}


\subsection{Simulated data}
 

Tensors were generated with combinations of $L_r$ and $R$ to test the BTDCORCONDIA as described in Section~\ref{corcond_simdata}.
For the case where $L_r$ varied between each component, the possible number of combinations of $L_r$ exponentially increases with the number of $R$
. Only the correct combinations of $L_r$, and $R$ resulted in 100\% BTDCORCONDIA. Moreover, adding 50 dB SNR to the tensor has little to no effect, with the core consistency value close to 99\%. 

\begin{figure}[!ht]
      \vspace{-1.0em}
      \centering
      \includegraphics[width=\linewidth]{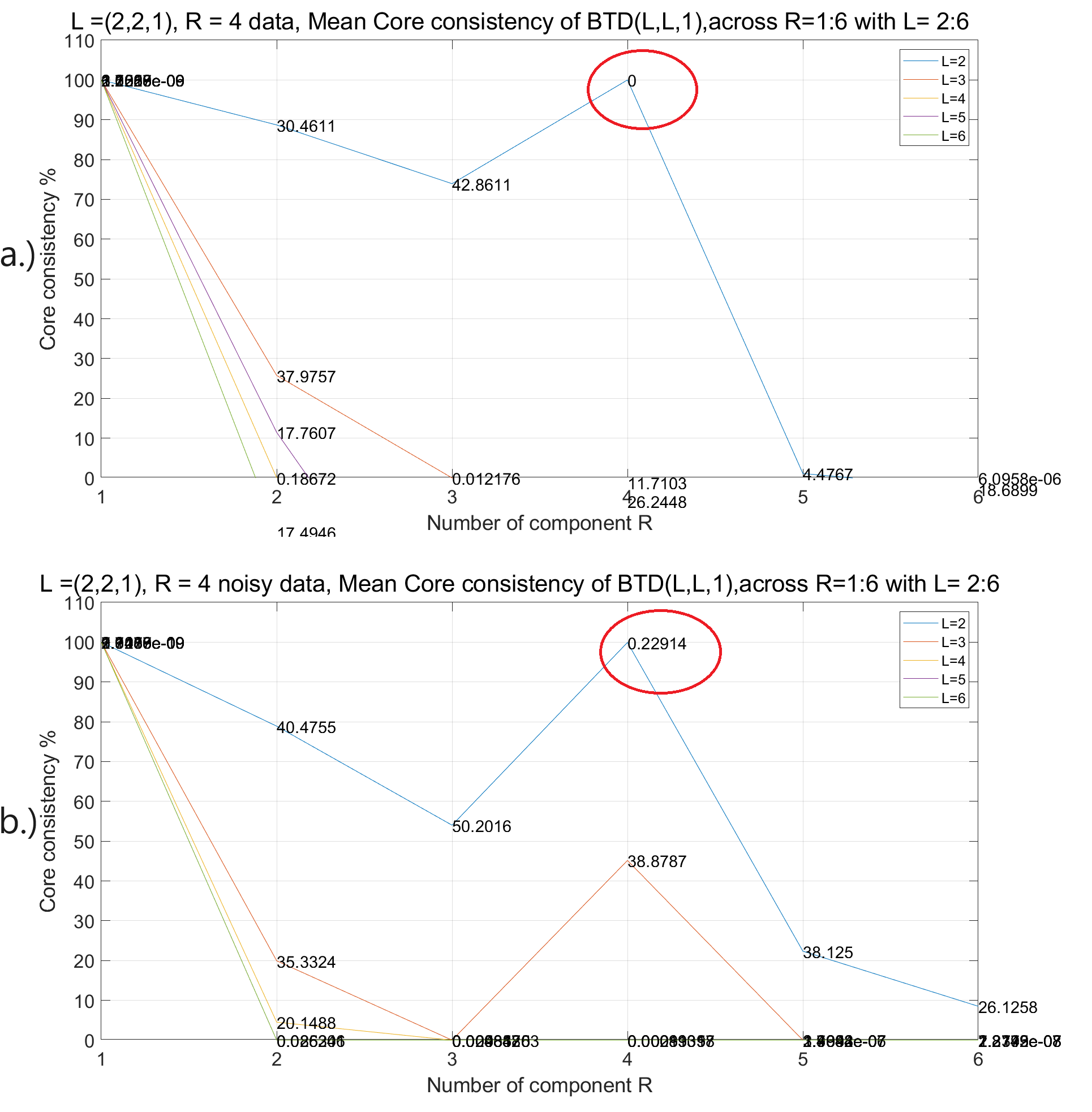}
      \caption{Mean core consistency of simulated tensor from $L_1  = \dots = L_r = 2$ and $R = 4$, were plotted with SD label on the graph. Figure a.) shows the results with no added noise, and b.) shows the result of the noisy tensor with SD slightly increased, as highlighted in the red circle.}
      \label{corcond_nonoise_noise}
      \vspace{-1.0em}
\end{figure}

Similarly, for $L_1 = L_2 = \dots = L_r$, BTDORCONDIA  always reached 100\% when the matching $L_r$ and $R$ are selected. However, with added noise, the mean percentage core consistency sometimes dropped. Fig.~\ref{corcond_nonoise_noise} compares simulated tensors with and without added noise when $L_1 = \dots = L_r = 2$ and $R = 4$. Both cases had a mean BTDCORCONDIA of almost 100\%, but with added noise, there are slight variations resulting in the SD value not being zero, as highlighted in the red circle. Nevertheless, the added noise usually reaches close to 99\% core consistency in most cases.

However, there were cases in which BTDCORCONDIA did not reach 100\%. For instance, this happened in some data generated with $R=2$ for $L_1 = \dots = L_r$. 
For investigation, 
those cases were re-run and manually inspected. 

We found that the core consistency is significantly reduced when feeding factors with a high cost to the BTDCORCONDIA model. This only happens when the BTD does not reach the global solution. By implementing additional refinements such as maximum interactions and ensuring the gevd initialisation, the issue of failed core consistency almost completely disappeared. This confirms that BTDCORCONDIA results in values of 100\% when the generated tensor follows exactly a BTD model.

\subsection{Exploration of the BTD indeterminacy} 

Small tensor sizes $10 \times 11 \times 12$ were tested and found that the core consistency achieved close to 100\% at lower $L_r$ and $R$. However, as the tensor was generated with $L_R \geq 4$ or $R \geq 4$, the BTDCORCONDIA missed the optimum point and could not find the $L_r$ and $R$ combination used, as expected. 
Similarly, when $L_r$ varied, the core consistency dropped when the total summation of $L_r$ exceeded ten ($\sum L_r \geq 10$). When $\sum L_r \geq 10$, the number of elements inside the core would be equal to or larger than the number of elements inside the tensor itself, which contradicts the nature of tensor operations as in \cite{bro2003new}. Additionally, the smallest side of the tensor should be $\geq \sum L_r$, so the gevd initialisation can work properly \cite{de2008decompositions2}. Therefore, tensor size $10 \times 11 \times 12$ was too small for BTD with high $L_r$ and $R$ to work properly. 


Next, we investigated how the arbitrary scale affects the core consistency. The random non-singular matrix \textbf{F} and its inversion were applied to the factors components. Then, the new applied factors were tested on BTDCORCONDIA, comparing it to the original factors and those with \textbf{F} and $\mathbf{F}^{-1}$ applied. The result showed no difference in BTDCORCONDIA between the original factors and those modified with \textbf{F} and $\mathbf{F}^{-1}$. 

\begin{figure}[!ht]
      \centering
      \includegraphics[width=\linewidth]{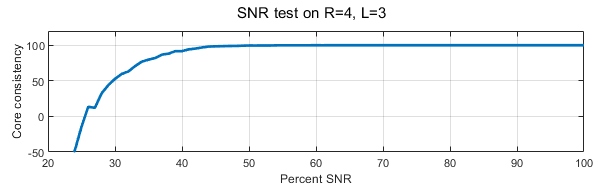}
      \caption{SNR plot for increasing noise level in factors decomposed from the ideal tensor generated from $R=4, L=3$. The core consistency remained high up to around 30 dB and started to drop at around 40 \% SNR.}
      \label{BTDcorcond_SNR}
      \vspace{-1.5em}
\end{figure}

For noise sensitivity, the ideal BTD tensor was decomposed, and increasing noise was added to their factors. Then, BTDCORCONDIA was computed for tensors with different levels of SNRs and plotted in Fig.~\ref{BTDcorcond_SNR}. The model tolerates some noise levels, as shown in Fig.~\ref{BTDcorcond_SNR}. The core consistency remained high up to around 30 dB SNR. It shows that as long as the components within the tensor interact in rank $L_r, L_r, 1$, the BTDCORCONDIA model can find the $R$ and $L_r$ combination even with some noise. 

\subsection{Real data}
Neu and CMI EEG tensors were decomposed by BTD with combinations of $R$ and $L_r$. Then, the EEG tensor and its factors were put into the BTDCORCONDIA model. Table~\ref{EEG_corcond} shows the top ten combinations of $R$ and $L_r$ with the highest core consistency. First, the combination that yields core consistency of 100 \% for both data are $R=2$, with $L1 = L2 = 1$, which is technically equal to CPD. The second highest for Neu tensor has 93.45\% core consistency from $R=2$, with $L_1 = 1$ and $L_2 = 3$. Meanwhile, the second highest core consistency for CMI tensor is still considered CPD with all $L_r$ equal to 1. The highest core consistency for BTD on the CMI tensor is from $R=2$, with $L_1 = 1$ and $L_2 = 3$, same as Neu tensor, but with only 40.27 \%, which suggests the BTD is unsuitable for the CMI tensor.

\begin{table}[]
\scriptsize
\caption{Ten highest core consistency from combinations of $R$ and $L_R$. The total components inside the bracket present $R$, while the value present each $L_R$. }
\label{EEG_corcond}
\begin{center}
\begin{tabular}{|l|l|l|l|}
\hline
\multicolumn{2}{c}{Neu EEG} & \multicolumn{2}{c}{CMI EEG} \\
\hline
$L_R$ & \%core & $L_R$ & \%core \\
\hline
 [1,1] & 100 & [1,1] & 100 \\
 
 [1,3] & 93.45 & [1,1,1] & 56.25 \\
 
 [1,6] & 90.93 & [1,3] & 40.27 \\
 
 [1,5] & 88.59 & [1,6] & 16.05 \\
 
 [1,4] & 79.85 & [2,6] & 9.52 \\
 
 [2,3] & 78.08 & [2,2] & 9.26 \\
 
 [2,4] & 58.79 & [2,5] & 9.23 \\
 
 [2,5] & 57.52 & [1,2] & 8.63 \\
 
 [2,6] & 40.39 & [1,3,6] & 5.36 \\
 
 [2,2] & 30.53 & [3,5] & 4.89 \\

\hline

\end{tabular}
\end{center}
\vspace{-1.5em}
\end{table}

After the combination for the Neu tensor was found, we inspected the BTD on the Neu tensor with $R=2$, with $L_1 = 1$ and $L_2 = 3$ (BTDCORCONDIA = 94.68\%).

\begin{figure}[!ht]
      \centering
      \includegraphics[width=\linewidth]{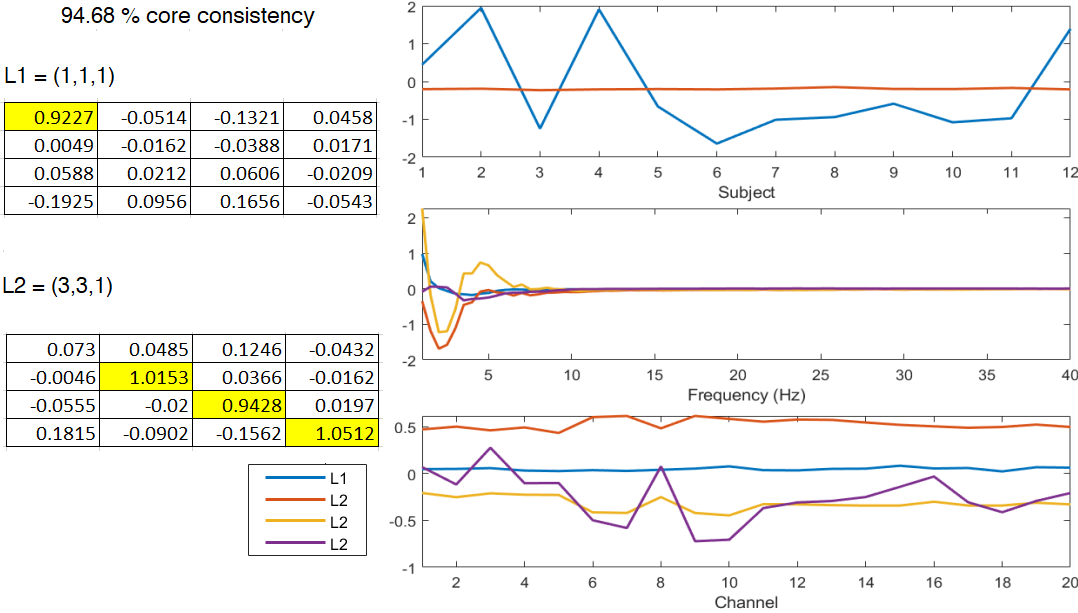}
      \caption{Core tensor and factor components plot from BTD on Neu EEG tensor with $R=2$, with $L_1 = 1$ and $L_2 = 3$. The blue component belongs to the first $L_1 = 1$ component, while red, yellow and purple belong to the second $L_2 = 3$ components. }
      \label{Neu_EEGcorcond_plot}
      \vspace{-1.5em}
\end{figure}

The decomposed core and factors are plotted in Fig.~\ref{Neu_EEGcorcond_plot}. Since $R=2$, the core tensor has two slices, with the diagonal value as close to 1, and the rest are close to 0, resembling the ideal core tensor. The factors are plotted on the side with a blue component from component 1, and the rest are from the second component with $L_2 = 3$. The plots show that all components were unique and can interact in the $(L_r, L_r, 1)$ ways. The EEG tensor was not imposed with constraints such as non-negativity. The two underlying components could refer to two groups of subjects residing within the Neu tensor. The blue frequency spectrum is only associated with the blue subject group and has frequency activity around $\leq 3$ Hz. The red component indicates that three frequency spectra (red, yellow, and purple) were associated or common occurrences within the red subject group and are related to different contributions of channel mode. The red and yellow spectrum has a similar activity range at $\leq 8$ Hz while the purple range around $\leq 6$ Hz with lower amplitude.


\section{Discussion}\label{btdcorcond_discussion}

The BTDCORCONDIA was tested on both simulated and real data tensors. When the correct combination of $R$ and $L_r$ is used, core consistency will always achieve $\approx 100 \%$. 



The real EEG tensors were tested. We found the signal in Neu EEG was compatible and appeared to interact in the multilinear way of BTD, with the core tensor resembling the ideal $\mathcal{II}$ core. Conversely, the CMI EEG did not follow the BTD but was compatible with CPD since all $L_r = 1$. 
Since BTDCORCONDIA is specifically designed for BTD, if the data within does not interact with multilinear trends of $(L_r, L_r, 1)$, there is a substantial likelihood that the core will not be set to identity during the iteration and will not yield a high percentage of consistency.

This model was built based on the same theory as CPD CORCONDIA \cite{bro2003new}. By including the $L_r =1$ across all $R$ during the search, the proposed BTDCORCONDIA can measure the data compatibility to not only BTD but CPD simultaneously. The model could supersede the traditional CPD CORCONDIA with the only drawback of computational expense to find the most appropriate model for the data.


It would be interesting to investigate and compare against a similar model with the known underlying components data, such as TuckCorCon \cite{kompany2012tucker}, where the core of restricted Tucker3 is measured. Even though the BTD can be viewed as a strict version of Tucker3, our approach to generating an ideal core and the main purpose differ. The work in \cite{kompany2012tucker} focused on measuring the compatibility between the sparseness of the core and the data to validate the hypothesised structure.

The experiment encountered computational time limitations while processing EEG tensors through a comprehensive grid search, considering all potential combinations of $R$ and $L_r$, particularly when $L_r$ varied. As $R$ increases, possible $L_r$ combinations grow exponentially. This posed an opportunity to improve the BTDCORCONDIA performance further to reduce the computation loads.

\section{Conclusion}

This paper demonstrated the principle of core consistency could be leveraged to estimate the rank for BTD $(L_r, L_r, 1)$. After evaluating the core consistency model on simulated tensors and real-world data, we concluded that BTDCORCONDIA is a promising approach to evaluate the model's component number and rank estimation, together with traditional visualisation and exploring the patterns of interactions between loadings of different modes. With the right settings, BTDCORCONDIA could test the data compatibility to both BTD and CPD simultaneously. However, the noise test suggests BTDCORCONDIA may not work well for high noise regimes.



\bibliographystyle{unsrt}
\bibliography{references}

\end{document}